\documentclass[conference]{IEEEtran}
\IEEEoverridecommandlockouts

\usepackage[pdftex]{graphicx}
\usepackage{amsmath}
\usepackage{amsfonts}
\usepackage{amssymb}
\usepackage{bm}
\usepackage{algorithmic}
\usepackage{array}
\usepackage[utf8]{inputenc}
\usepackage[T1]{fontenc}
\usepackage[caption=false,font=normalsize,labelfont=sf,textfont=sf]{subfig}
\usepackage{fixltx2e}
\usepackage{stfloats}
\usepackage{url}
\usepackage{siunitx}
\usepackage[english, onelanguage,ruled,vlined, longend, titlenotnumbered]{algorithm2e}
\usepackage{pgfplots}
\usepackage{tikz}
\usepackage[english, plain]{fancyref}
\usetikzlibrary[patterns]
\newcommand{\bfA}{\mathbf{A}}

\newcommand{\tbfA}{\tilde{\mathbf{A}}}

\newcommand{\btbfA}{\bar{\tilde{\mathbf{A}}}}
\newcommand{\bfa}{\mathbf{a}}

\newcommand{\bfC}{\mathbf{C}}

\newcommand{\bbC}{\mathbb{C}}

\newcommand{\bfd}{\mathbf{d}}
\newcommand{\hbfd}{\hat{\mathbf{d}}}

\newcommand{\bfD}{\mathbf{D}}

\newcommand{\calE}{\mathcal{E}}

\newcommand{\sfF}{\mathsf{F}}
\newcommand{\bfg}{\mathbf{g}}

\newcommand{\bbfg}{\bar{\mathbf{g}}}
\newcommand{\hbfg}{\hat{\mathbf{g}}}
\newcommand{\bfG}{\mathbf{G}}
\newcommand{\hbfG}{\hat{\mathbf{G}}}
\newcommand{\bbfG}{\bar{\mathbf{G}}}
\newcommand{\calH}{\mathcal{H}}
\newcommand{\sfH}{\mathsf{H}}
\newcommand{\bfI}{\mathbf{I}}

\newcommand{\bfn}{\mathbf{n}}
\newcommand{\rmn}{\mathrm{n}}

\newcommand{\bfN}{\mathbf{N}}

\newcommand{\bfp}{\mathbf{p}}
\newcommand{\hbfp}{\hat{\mathbf{p}}}

\newcommand{\hbfr}{\hat{\mathbf{r}}}
\newcommand{\chbfr}{\check{\hat{\mathbf{r}}}}
\newcommand{\bfR}{\mathbf{R}}
\newcommand{\bbR}{\mathbb{R}}
\newcommand{\hbfR}{\hat{\mathbf{R}}}

\newcommand{\bfs}{\mathbf{s}}

\newcommand{\calS}{\mathcal{S}}

\newcommand{\sfT}{\mathsf{T}}

\newcommand{\bfx}{\mathbf{x}}

\newcommand{\bfW}{\mathbf{W}}

\newcommand{\rmw}{\mathrm{w}}

\newcommand{\bfz}{\mathbf{z}}

\newcommand{\boGamma}{\boldsymbol{\Gamma}}

\newcommand{\boomega}{\boldsymbol{\omega}}
\newcommand{\boOmega}{\boldsymbol{\Omega}}

\newcommand{\bosigma}{\boldsymbol{\sigma}}
\newcommand{\boSigma}{\boldsymbol{\Sigma}}
\newcommand{\cbosigma}{\check{\boldsymbol{\sigma}}}
\newcommand{\tbosigma}{\tilde{\boldsymbol{\sigma}}}
\newcommand{\hbosigma}{\hat{\boldsymbol{\sigma}}}
\newcommand{\htbosigma}{\hat{\tilde{\boldsymbol{\sigma}}}}
\newcommand{\tboSigma}{\tilde{\boldsymbol{\Sigma}}}

\newcommand{\boxi}{\boldsymbol{\xi}}
\newcommand{\boXi}{\boldsymbol{\Xi}}

\newcommand{\bfRzero}{\mathbf{R}^{\TRUE}}
\newcommand{\hbfRzero}{\hat{\mathbf{R}}^{\TRUE}}

\DeclareMathOperator{\diag}{diag}

\DeclareMathOperator{\vect}{vec}
\DeclareMathOperator{\vectdiag}{vecdiag}
\DeclareMathOperator*{\argmin}{arg\,min}

\newcommand{\bfun}{\mathbf{1}}
\newcommand{\bfzero}{\mathbf{0}}
\newcommand{\IN}{\text{\tiny K}}
\newcommand{\OUT}{\text{\tiny U}}
\newcommand{\TRUE}{\text{\tiny K}}

\usetikzlibrary{plotmarks}
\usepackage[nospace]{cite}
\usepackage{eso-pic}
\begin{document}

\title{Calibration of Radio Interferometers Using a Sparse DoA Estimation Framework}

\author{\IEEEauthorblockN{M. \textsc{Brossard}\IEEEauthorrefmark{4},
M. N. \textsc{El Korso}\IEEEauthorrefmark{1},
M. \textsc{Pesavento}\IEEEauthorrefmark{2}, 
R. \textsc{Boyer}\IEEEauthorrefmark{3} and P. \textsc{Larzabal}\IEEEauthorrefmark{6} 
}
\IEEEauthorblockA{\IEEEauthorrefmark{4}\IEEEauthorrefmark{6}SATIE, UMR 8029, École Normale Supérieure de Cachan, Cachan, France}
\IEEEauthorblockA{\IEEEauthorrefmark{1}Université Paris Ouest Nanterre La Défense, IUT de Ville d’Avray, LEME EA 4416, France}
\IEEEauthorblockA{\IEEEauthorrefmark{4}\IEEEauthorrefmark{2}Communication Systems Group, Technische Universität, Darmstadt, Germany}
\IEEEauthorblockA{\IEEEauthorrefmark{3}L2S, UMR 8506, Université Paris Sud, Gif-sur-Yvette, France}

\thanks{This work was supported by the following projects: MAGELLAN
(ANR-14-CE23-0004-01), MI-CNRS TITAN and ICode blanc.}
\markboth{Journal of Quantum Telecommunications, ̃Vol
. ̃1, No. ̃1, ̃January ̃2025}{Shell \MakeLowercase{\text
it{et al.}}: A Novel Tin Can Link}
}
\maketitle

\begin{abstract}
The calibration of modern radio interferometers is a significant challenge, specifically at low frequencies. In this perspective, we propose a novel iterative calibration algorithm, which employs the popular sparse representation framework, in the regime where the propagation conditions shift dissimilarly the directions of the sources. More precisely, our algorithm is designed to estimate the apparent directions of the calibration sources, their powers, the directional and undirectional complex gains of the array elements and their noise powers, with a reasonable computational complexity. Numerical simulations reveal that the proposed scheme is statistically efficient at low SNR and even with additional non-calibration sources at unknown directions.
\end{abstract}


\section{Introduction}

The calibration is a salient challenge for the new generation of radio interferometers \cite{Wij2010}, such as the LOw Frequency ARray (LOFAR) \cite{Haa2013} or the Square Kilometer Array (SKA) \cite{Dew2009}. These instruments consist in large sensor arrays for which calibration is essential to produce accurate images (\SI{>e8}{pixels}) with high dynamic range (\SI{60}{dB}). Furthermore, the huge number of array elements imposes the need for designing computationally efficient algorithms.

For such radio interferometers, additional difficulties arise at low frequencies (\SI{<300}{MHz}), where the ionosphere causes phase delays which scale with the wavelength \cite{Vanthe, Int2009}. In this paper, we focus on the regime where all lines of sight toward a source in the sky cross the same ionospheric layer, where the thickness of the ionosphere can be direction dependent \cite{Lon2004}. In this regime, the ionospheric phase delays are commonly modeled as a linear function of the distance between the so-called piercing points \cite{Cot2004, Ven2013}. As a consequence, it modifies the geometric delays and introduces direction dependent angular-shifts for the source directions. By estimating the calibrator shifts (i.e., the difference between the true calibrator directions, known from tables \cite{Ben1962}, and their apparent directions, estimated from the observations of the radio interferometers), an interpolation method can be efficiently applied in order to obtain a phase screen model, that 
estimates the ionospheric delays over the entire Field-of-View \cite{Cot2004}. In addition to the phase screen reconstruction step, calibration usually involves the estimation of the complex undirectional gains of the antennas, their directional gains toward each calibrator and their noise powers\cite{Wijthe}.

To solve this calibration problem, the a priori knowledge of some calibration sources is required, i.e., their true/nominal directions and powers without the effects of the ionosphere and antenna imperfections \cite{Ven2013}. Based on this knowledge, state-of-the-art calibration algorithms to estimate the aforementioned parameters are mostly of iterative nature \cite{Wijthe, Ven2013, Wij2010}. As an example, the (Weighted) Alternating Least Squares approach has been adapted for LOFAR calibration \cite{Wijthe}, in which closed form expressions have been obtained for the undirectional antenna gains, the source powers and the sensor noise powers. Nevertheless, regarding to the Direction-of-Arrival (DoA) estimation, no closed form expression can be obtained and  classical subspace methods, such as MUSIC \cite{Sch1986}, have to be applied. A major drawback of these methods is that subspace techniques are not efficient in low Signal-to-Noise-Radio (SNR) scenarios and require the exact number of sources in the 
scene.

As an alternative approach, recently, sparse reconstruction methods came into focus of DoA estimation for fully calibrated arrays \cite{Mal2005} as well as for partially calibrated arrays \cite{Ste2014}. They exhibit the super-resolution property, robustness and computational efficiency, without the aforementioned drawbacks of subspace-based methods \cite{Mal2005}. Sparse representation methods have been successfully applied in radio astronomy for imaging \cite{Wia2009}, but, to the best of our knowledge, such methods have never been applied for this calibration problem.

In this paper, we focus on the calibration of a sensor array, involving its individual antennas and propagation disturbances. In addition, we assume that the sensor array has an arbitrary geometry, identical elements and is simultaneously excited by inaccurately known calibration sources and unknown non-calibration sources. We consider these non-calibration sources as outliers, i.e., as an additional noise term in the calibration step. From the calibration perspective, we propose a novel iterative scheme, which successively estimates the undirectional antenna gains along with the calibrator and noise parameters, to minimize a proper weighting cost function. The calibrator  parameter estimation relies on the popular sparse representation framework and the sensor noise power estimation considers the presence of non-calibration sources (a.k.a. outliers in our calibration procedure), leading to a robust and computationally efficient algorithm in low SNR scenarios.

In the following, $\bar{(.)}, (.)^{\sfT}, (.)^{\sfH}, (.)^{\dagger}, (.)^{\odot \alpha}$ and $[.]_{n}$ denote, respectively, conjugation, transpose, Hermitian transpose, pseudo-inverse, element-wise raising to $\alpha$ and the $n$-th element of a vector. The expectation operator is $\calE\{.\}, \circ$ denotes the Khatri-Rao product, $\exp(.)$ and $\odot$ represent the element-wise exponential function and multiplication (Hadamard product), respectively. The operator $\diag(.)$ converts a vector to a diagonal matrix with the vector aligned on the main diagonal, whereas $\vectdiag(.)$ produces a vector from the main diagonal of its entry and $\vect(.)$ converts a matrix to a vector by stacking the columns of its entry. The functions $\left\| . \right\|_0, \left\|.\right\|_2$ and $\left\|.\right\|_\sfF$ refer to $l_0$ norm, i.e., the number of non-zero elements of its entry, the $l_2$ and Frobenius norms, respectively. Finally, $\bfx \succeq0$ means that each element in its vector $\bfx$ is non-negative.

\section{Data model and problem statement}
Let us consider an array of $P$ elements, with known locations, each detonated by the Cartesian coordinates $\boxi_p = \left[x_p, y_p, z_p \right]^{\sfT}$ for $p = \allowbreak 1,\ldots,P$, that we stack in $\boXi = \left[\boxi_1,\ldots,\boxi_P \right]^{\sfT} \in \bbR^{P \times 3}$. This array is exposed to $Q$ known strong calibration sources and additional $Q^{\OUT}$ unknown weak non-calibration sources, with known true $\bfD^{\IN} = \left[ \bfd^{\IN}_1 , \ldots , \bfd^{\IN}_{Q} \right]  \in \bbR^{3 \times Q}$  and unknown $\bfD^{\OUT} = \left[ \bfd^{\OUT}_1 , \ldots , \bfd_{Q^{\OUT}}^{\OUT} \right] \in \bbR^{3 \times Q^{\OUT}}$ spatial coordinates, respectively, in which each direction $\bfd = \left[l,m,n\right]^{\sfT}$  can be uniquely described by a couple $(l,m)$, since $n = \allowbreak \sqrt{1-l^2-m^2}$ \cite{Ven2013}. The ionosphere introduces an unknown angular-shift for each source direction \cite{Cot2004, Wij2010}, and consequently, we distinguish between the unknown \emph{apparent} directions for 
the 
calibrators, denoted by $\bfD = \left[ \bfd_1 ,\ldots , \bfd_{Q} \right]$, and their \emph{true} directions $\bfD^{\IN}$, i.e., without the propagation disturbances. 

Under the narrowband assumption, the steering vector $\bfa(\bfd)$ toward the direction $\bfd$ is given by
\begin{equation}
\bfa(\bfd) = \bfa(l,m) =  \frac{1}{\sqrt{P}} \exp \left( - \mathrm{j} \frac{2\pi}{\lambda} \boXi \bfd \right) \in \bbC^{P} \text{,}
\end{equation}
where $\lambda$ denotes the wavelength. For the calibration source signals, we consider the steering matrix
\begin{equation}
 \bfA = \frac{1}{\sqrt{P}} \exp \left( -\mathrm{j} \frac{2\pi}{\lambda} \boXi \bfD \right) \in \bbC^{P \times Q} \text{,}
\end{equation}
which contains the calibrator steering vectors. We define $\bfA^{\OUT}$, w.r.t. $\bfD^{\OUT}$, correspondingly for the non-calibration sources.

As in \cite{Wijthe}, we assume that all antennas have identical directional responses. Their directional gain responses (and propagation losses) can be modeled by two diagonal matrices $\boGamma \in \allowbreak \bbC^{Q \times Q}$ and $\boGamma^{\OUT} \in \allowbreak \bbC^{Q^{\OUT} \times Q^{\OUT}}$ toward the calibration and non-calibration sources, respectively.

The received signals from each antenna are stacked, for the $n$-th observation, into the vector
\begin{equation}
 \bfx(n) = \bfG \left( \bfA\boGamma\bfs(n)  + \bfA^{\OUT}\boGamma^{\OUT}\bfs^{\OUT}(n) \right) + \bfn(n) \label{eq:signal}\text{,}
\end{equation}
where $\bfG = \diag(\bfg) \allowbreak \in \bbC^{P \times P}$ models the undirectional antenna gains, $\bfs(n) \in \allowbreak \bbC^{Q}$ and  $\bfs^{\OUT}(n) \in \allowbreak \bbC^{Q^{\OUT}}$ represent, respectively, the i.i.d. calibrator and non-calibrator signals and the vector $\bfn(n) \sim \mathcal{CN}(\bfzero,\boSigma^{\rmn})$ denotes the i.i.d. noise, in the $n$-th observation \cite{Ven2013}. Consequently, the covariance matrix $\bfR = \allowbreak \calE \left\lbrace \bfx \bfx^{\sfH} \right\rbrace  $ of the observations is given by  \setlength{\arraycolsep}{0.0em}
\begin{equation}
\begin{aligned}
  \bfR &= \bfG\bfA \boGamma\boSigma^{\IN}\boGamma^{\sfH}\bfA^{\sfH}\bfG^{\sfH} +  \\
 & \bfG\bfA^{\OUT}\boGamma^{\OUT}\boSigma^{\OUT}\boGamma^{\OUT\sfH}\bfA^{\OUT\sfH}\bfG^{\sfH} + \boSigma^{\rmn} \text{,}
\end{aligned}
\end{equation}
where $\boSigma^{\IN} \in \bbR^{Q \times Q}, \boSigma^{\OUT} \in \bbR^{Q^{\OUT} \times Q^{\OUT}}$ and $\boSigma^{\rmn} = \diag\left(\bosigma^{\rmn}\right) \in \bbR^{P \times P}$ denote, respectively, the diagonal covariance matrix for the calibrators, non-calibration sources and sensor noises. \setlength{\arraycolsep}{5pt}

As $\boGamma$ and $\boGamma^{\OUT}$ are diagonal matrices, we define in the sequel the diagonal matrix that contain the apparent calibrator powers as $\boSigma = \allowbreak \boGamma\boSigma^{\IN}\boGamma^{\sfH} = \allowbreak \diag \left(\bosigma \right) \in \allowbreak \bbR^{Q \times Q}$. Since only $\boSigma^{\IN}$ is assumed to be known, $\boSigma$ is generally unknown and $\boGamma$ can be deduced from it. We further introduce the unknown covariance matrix for the non-calibration sources, $\bfR^{\OUT} = \bfG\bfA^{\OUT}\boGamma^{\OUT}\boSigma^{\OUT}\boGamma^{\OUT\sfH}\bfA^{\OUT\sfH}\bfG^{\sfH}$, and rewrite the covariance matrix model \cite{Wij2010} as
\begin{equation}
 \bfR  = \bfG\bfA\boSigma\bfA^{\sfH}\bfG^{\sfH} +\bfR^{\OUT} + \boSigma^{\rmn} \text{.}
\label{eq:cov}
\end{equation}

We then formulate the calibration problem as the estimation of the parameter vector $\bfp = \left[\bfg^{\sfT}, \bfd^{\sfT}_{1}, \ldots, \bfd^{\sfT}_{Q}, \bosigma^{\sfT}, \bosigma^{\rmn \sfT} \right]^{\sfT}$, from the sample covariance matrix $\hbfR = \frac{1}{N} \sum_{n=1}^N \bfx(n) \bfx^{\sfH}(n)$. Note that the estimation of the unknown matrix $\bfR^{\OUT}$ represents the imaging step \cite{Wijthe} which is beyond the scope of the paper. The imaging step is done usually as a separate step after the calibration \cite{Wia2009}. The main reason is that the calibration step is usually based on a source point model (unlike the imaging step) with a known number of calibrators, whereas, the effect of the weakest (non-calibration) sources, with an unknown source number, can be assumed absorbed by the noise component.

\begin{algorithm}[b]
 \KwIn{sample covariance matrix $\hbfR$\;}
 \textbf{Init:} set the iteration counter $i = 0$, $\bfg = \bfg^{[0]}, \bfD=\bfD^{\IN}, \bosigma= \diag \left( \boSigma^{\IN} \right), \boOmega = \bfun_{P \times P}$\;
 \While{$\left\| \bfp^{[i-1]}-\bfp^{[i]} \right\|_2 \ge \left\| \bfp^{[i]} \right\|_2 \epsilon_{\bfp}$}{
  \nl $i = i+1$\;
  \nl Estimate $\bfg^{[i]}$ \label{lst:line:g} with Algorithm 2\;
  \nl Estimate $\bfD^{[i]}, \bosigma^{[i]}$ and $\bosigma^{\rmn [i]}$ \label{lst:line:doa} with Algorithm 3\;
  \nl Update $\boOmega = \left( \bosigma^{\rmn [i]} \bosigma^{\rmn [i] \sfT} \right)^{\odot -\frac{1}{2}}$\;}
 \KwOut{$\hbfp = \left[\bfg^{[i]\sfT}, \bfd^{[i]\sfT}_{1}, \ldots, \bfd^{[i]\sfT}_{Q}, \bosigma^{[i]\sfT}, \bosigma^{\rmn[i]\sfT} \right]^{\sfT}$\;}

 \caption[]{Iterative and Sparsity Based Calibration \\ Algorithm (ISBCA)\label{algo:proposed}}
\end{algorithm}

\renewcommand{\thealgocf}{}

In order to overcome the scaling ambiguities in model (\ref{eq:cov}), we consider the following commonly used assumptions in radio astronomy \cite{Wijthe, Ven2013}: i) to solve the phase ambiguity of $\bfg$, we take its first element as the phase reference; ii) $\bfg$ and $\bosigma$ share a common scalar factor and when solving for the calibrator directions, a single rotation of all steering vectors can be compensated by the undirectional gain phase solution. We therefore consider an additional known source to remove both ambiguities, by fixing its direction $\bfd_0$ and its apparant power $\sigma_0$.

\section{Calibration scheme}
\label{sec:algo}
\subsection{Overview of the Proposed Algorithm}

A statistically efficient estimator of the model parameters can be obtained via the Maximum-Likehood formulation, but it appears intractable in practice. However, with a large number of samples, statistically efficient estimations can be reached using the Weighting Least Squares approach. Consequently, we define the following cost function to minimize: $\kappa(\bfp) = \left\| \bfW^{-\frac{1}{2}} \left( \bfR(\bfp) - \hbfR \right) \bfW^{-\frac{1}{2}} \right\|_{\sfF}^2$, with $\bfR(\bfp) = \bfG\bfA\boSigma\bfA^{\sfH}\bfG^{\sfH} + \boSigma^{\rmn}$ denoting the covariance matrix in the absence of the non-calibration sources, and $\bfW$ is the weighting matrix. The optimal weighting matrix for  Gaussian noise is $\bfW = \bfR$ \cite{Wijthe}, which is generally unknown. In radio astronomy, sources are typically much weaker than the noise \cite{Sal2014}, so the covariance matrix can be approximated by $\bfR \approx \boSigma^{\rmn}$. Since the array consists of identical antennas and mutual coupling is negligible, it 
is commonly 
assumed that $\boSigma^{\rmn} = \allowbreak \diag \left(\bosigma^{\rmn} \right) \approx  \allowbreak \sigma^{\rmn} \bfI$. Consequently, we consider $\bfW = \bfI$ as an initial step and refine it with $\bfW = \boSigma^{\rmn}$ once we obtain an estimate of $\boSigma^{\rmn}$. Since $\boSigma^{\rmn}$ is diagonal, we rewrite the cost function as
\begin{equation}
\kappa(\bfp) = \left\| \left( \bfR(\bfp) - \hbfR \right) \odot \boOmega \right\|_{\sfF}^2  \label{eq:kappa:2} \text{,}
\end{equation}
with $\boOmega = \left( \bosigma^{\rmn} \bosigma^{\rmn \sfT} \right)^{\odot -\frac{1}{2}}$.

We aim at minimizing the cost function $\kappa(\bfp)$ in an iterative manner. We first minimize $\kappa(\bfp)$ w.r.t. $\bfg$, with the remaining parameters in $\bfp$ fixed as described in the Subsection III-B. In a sequential step, we minimize (\ref{eq:kappa:2}) w.r.t. the variables $\bosigma, \bfd_1, \ldots, \bfd_{Q}$ and $\bosigma^{\rmn}$ for fixed  $\bfg$, by using a sparse representation approach as described in the Subsection III-C. The overall procedure, referred to as the Iterative and Sparsity Based Calibration Algorithm (ISBCA), is presented in Algorithm \ref{algo:proposed}. The algorithm is initialized with the true/nominal calibrator parameters and an initial guess for the undirectional gains or the unit sensor gain. In the following, we detail the two major alternating optimization steps of the ISBCA.

\begin{algorithm}[b]
\SetAlgoRefName{1.2}
 \KwIn{sample covariance matrix $\hbfR$\;}
 \textbf{Init:} set the iteration counter $k = 0$, $\bfg = \bfg^{[i-1]}, \bfRzero = \bfA^{[i-1]} \boSigma^{[i-1]} \bfA^{[i-1]\sfH}$\;
  \While{$\left\| \bfg^{[k-1]}-\bfg^{[k]}\right\|_2 \ge \left\| \bfg^{[k]} \right\|_2 \epsilon_{\bfg}$}{
   \nl $k = k+1$ \;
   \For{$p = 1, \ldots, P$}{
     \nl$\hbfr_p = \calS_p\left(\hbfR\right)$ \;
     \nl$\bfz = \calS_p \left(\bfRzero \bbfG \right)$ \;
     \nl$\bfz_{\rmw} = \bfz \odot \calS_p \left(\boOmega\right)$ \;
    \nl$[\bfg]_{p}^{[k]} = \frac{\bfz_{\rmw}^{\sfH}  \hbfr_p}{\bfz_{\rmw}^{\sfH} \bfz} $ \;
    }
  }
 \KwOut{$\hbfg = \bfg^{[k]}$\;}
 \caption{undirectional antenna gain estimation\label{alg:gain_estimation}}
\end{algorithm}

\subsection{Undirectional Antenna Gain Estimation (Algorithm 1.2)}
In this subsection, we describe Algorithm 1.2 of the ISBCA. We follow the same iterative approach as discussed in \cite{Sal2014}, that  we adapt for our cost function (\ref{eq:kappa:2}) that we optimize w.r.t. $\bfg$ for the remaining parameter in $\bfp$ fixed.

Toward this aim, we consider $\bfg$ and $\bbfg$ as two independent variables. We first regard $\bbfg$ as fixed and minimize (\ref{eq:kappa:2}) w.r.t.  $\bfg$, only, and without considering the diagonal elements in the cost function (\ref{eq:kappa:2}) that contain the unknown noise variances $\bosigma^{\rmn}$. In this case, the cost function becomes separable w.r.t. the elements of $\bfg$, hence,
\begin{equation}
 \kappa(\bfg) = \sum_{p=1}^{P} \kappa^{p}([\bfg]_{p}) \text{,} \label{eq:kappa:n}
\end{equation}
where $\kappa^{p}([\bfg]_{p})$ corresponds to the cost function for the $p$-th row of $\bfR$, which depends only on $[\bfg]_{p}$ since the remaining parameters are considered as fixed in this step. Let us define the operator $\calS_p(.)$, that converts to a vector the $p$-th row of a matrix and removes the $p$-th element of this selected vector. Further, define the vector $\hbfr_p = \calS_p\left(\hbfR\right)$ and the weighting vector $\boomega = \calS_p\left(\boOmega \right)$.  We can thus write $\kappa^{p}([\bfg]_{p})$ in (\ref{eq:kappa:n}) as
\begin{equation}
 \kappa^{p}([\bfg]_{p}) = \left\| \left(\hbfr_p - \bfz [\bfg]_{p} \right) \odot \boomega \right\|_2^2\text{,}\label{eq:kappa:3}
\end{equation}
in which $\bfz = \calS_p \left(\bfRzero \bbfG \right)$ and where $\bfRzero = \bfA \boSigma \bfA^{\sfH}$ represents the calibrator sky model. Since $\kappa^{p}([\bfg]_{p})$ is a Least Squares function of $[\bfg]_{p}$, by using standard inversion techniques and introducing $\bfz_{\rmw} = \bfz \odot \calS_p \left(\boOmega\right)$, its minimizer is given by $[\hbfg]_p = \frac{\bfz_{\rmw}^{\sfH}  \hbfr_p}{\bfz_{\rmw}^{\sfH} \bfz}$. After this, we directly update $[\bbfg]_p$ and proceed in the same manner with the remaining parameters in $\bfg$. This procedure, summarized  in Algorithm 1.2, is repeated until convergence.

\subsection{Calibrator Parameter and Noise Power Estimation \newline ~(Algorithm 1.3)}
In this subsection, we describe Algorithm 1.3 of the ISBCA for optimizing (\ref{eq:kappa:2}) w.r.t. the calibrator parameters and noise powers, that is based mainly on the popular sparse representation framework.

Assuming that the calibration sources are well separated, which is common in radio astronomy \cite{Wijthe, Ven2013}, we consider in the sequel that: i) each apparent calibration source lies in an sector of displacements around its nominal location; ii) the displacement sectors of different calibration sources are not overlapping; iii) each following dictionary shall represent the displacement set corresponding to its source.

Let us define $Q$ dictionaries of steering vectors $\tbfA_q$, for $q=1,\ldots, Q$, as
\setlength{\arraycolsep}{0.0em}
\begin{equation}
\begin{aligned}
 \tbfA_q =&~\Big[ \bfa(l_{q,1}, m_{q,1}), \ldots ,\bfa(l_{q,1}, m_{q,N_q^{m}}), \bfa(l_{q,2}, m_{q,1}),  \\
  & \ldots, \bfa(l_{q,N_q^{l}}, m_{q,N_q^{m}}) \Big] \in \bbC^{P \times N_q^{l}N_q^{m}} \label{eq:fat_a} \text{,}
\end{aligned}
\end{equation}
which contain  $N_q^{l}N_q^{m}$ steering vectors, centered around the true/nominal direction of the $q$-th calibrator, namely $\bfd^{\IN}_q$, with resolution $\left( \Delta l_q,\Delta m_q \right)$ and $N_q^{l} \gg 1, N_q^{m} \gg 1$. These $Q$ dictionary steering matrices are gathered in \setlength{\arraycolsep}{5pt}
\begin{equation}
 \tbfA = \left[ \tbfA_1, \ldots ,\tbfA_{Q} \right] \in \bbC^{P \times N_g}\text{,}
\end{equation}
with $N_g = \sum_{q=1}^{Q} N_q^{l}N_q^{m} $ denoting the total number of directions on the grid. We define the sparse calibrator power vector as
\begin{equation}
 \tbosigma = \left[ \tbosigma_1^{\sfT}, \ldots ,\tbosigma_{Q}^{\sfT} \right]^{\sfT} \in \bbR^{N_g} \text{,}
\end{equation}
which contains the powers of all calibrators. Due to the previous assumption of non-overlapping displacement sectors, we postulate that $\tbosigma_q$ is exactly $1$-sparse, i.e., $\left\| \tbosigma_q \right\|_0=1$.

\begin{algorithm}[ht]
\SetAlgoRefName{1.3}
 \KwIn{sample covariance matrix $\hbfR$\;}
 \textbf{Init:} set the iteration counter $k = 0$, $\bfg = \bfg^{[i-1]}, \bosigma = \bosigma^{[i-1]}, \bfD = \bfD^{[i-1]}$\;
 \While{$\left\| \tbosigma^{[k-1]}- \tbosigma^{[k]} \right\|_2 \ge \left\| \tbosigma^{[k]} \right\|_2 \epsilon_{\tbosigma}$}{
 \nl $k = k+1$\;
 \For{$q=1,\ldots,Q$}{
  \nl Calculate the residual vector for the $q$-th source: $\chbfr_{q} = \chbfr - \sum_{q' = 1}^{q-1} \check{\bfC}_{q'} \tbosigma_{q'}^{[k]} - \sum_{q'=q+1}^{Q} \check{\bfC}_{q'} \tbosigma_{q'}^{[k-1]}$\;
  \nl Calculate the stepsize $\mu_{q}^{k}$ as in \cite{Blu2010}\;
  \nl $\tbosigma_q^{[k]} =  \calH_1 \left( \tbosigma_q^{[k-1]} + \mu_q^k \check{\bfC}^{\dagger}_{q} \left(\check{\hbfr}_{q} - \check{\bfC}_{q} \tbosigma_q^{[k-1]} \right) \right)$ \label{lst:line:sig}\;}}
 \nl Estimate $\bosigma^{\rmn}$ with (\ref{eq:sigman:2})\;
 \caption{calibrator parameter and noise power \newline estimation\label{alg:iht}}
 \KwOut{$\htbosigma = \tbosigma^{[k]}$, that leads to $\hbosigma, \hbfd_{1}, \ldots, \hbfd_{Q}$ and $\hbosigma^{\rmn}$ \;}
\end{algorithm}

Using (\ref{eq:cov}), the covariance model can be rewritten as
\begin{equation}
 \bfR = \bfG \tbfA \tboSigma \tbfA^{\sfH} \bfG^{\sfH} + \bfR^{\OUT} + \boSigma^{\rmn} \text{,}
\end{equation}
in which $\tboSigma = \diag(\tbosigma)$. Let us then define
\setlength{\arraycolsep}{0.0em}
\begin{align}
 \bfC_q &=  \left( \boSigma^{\rmn -\frac{1}{2}} \bbfG \btbfA_q \right) \circ \left( \boSigma^{\rmn -\frac{1}{2}} \bfG \tbfA_q \right), q=1,\ldots, Q\\
 \bfC &= \left[\bfC_1, \ldots, \bfC_{Q} \right] = \left( \boSigma^{\rmn -\frac{1}{2}} \bbfG \btbfA \right) \circ \left( \boSigma^{\rmn -\frac{1}{2}} \bfG \tbfA \right)\\
 \bfN &= \boSigma^{\rmn -\frac{1}{2}} \circ \boSigma^{\rmn -\frac{1}{2}},  
 \hbfr = \vect(\hbfR \odot \boOmega)\text{.}
\end{align}
Thus, we formulate the minimization problem of (\ref{eq:kappa:2}) as \setlength{\arraycolsep}{5pt}
\begin{equation}
\begin{aligned}
&\htbosigma, \hbosigma^{\rmn} = \argmin_{\tbosigma, \bosigma^{\rmn}} \left\| \hbfr - \bfC \tbosigma - \bfN \bosigma^{\rmn} \right\|^2_2 \\
  &\text{subject to~} \tbosigma \succeq 0, \bosigma^{\rmn} \succeq 0, \left\| \tbosigma_q \right\|_0 = 1, q=1,\ldots, Q \text{,} \label{eq:min1}
\end{aligned}
\end{equation}
in which we make use of the fact that the $Q$ displacement sectors are not overlapping.

To consider the $l_0$ constraints in (\ref{eq:min1}), which are generally difficult, we choose the Iterative Hard Thresholding scheme of \cite{Blu2008}. This greedy algorithm is based on a projected gradient descend direction algorithm and offers strong theoretical guarantees that have already been employed in the DoA estimation context \cite{Oll2014}. Particularly, when the grid is fine and the columns of $\tbfA_q$ are strongly similar, we can guarantee that each $\tbosigma_q$ obtained from (\ref{eq:min1}) is exactly $1$-sparse. Thus, using the Coordinate Descent algorithm \cite{Fri2007} to minimize (\ref{eq:min1}), we obtain an analytic solution for each sub-problem and the sparsity of the desired minimizer $\tbosigma$ reduces the computational complexity\cite{Fri2007}. Each step involves the hard thresholding operator $\calH_s(.)$, that keeps the $s$-largest components of a vector and sets the remaining entries equal to zero, thus, it automatically satisfies both constraints of sparsity and positivity. We 
allow a step size factor $\mu^k_q$ that depends on $\tbosigma_q$ and the $k$-th iteration, hence considering the Normalized Iterative Hard Thresholding procedure of \cite{Blu2010}, where the choice of $\mu^k_q$ assures convergence toward a local minimum.

Since the $p$-th element of $\bosigma^{\rmn}$, $[\bosigma^{\rmn}]_{p}$, is only present in the $p$-th diagonal term of $\bfR$, ignoring this term does not affect the estimation of $\tbosigma$. Thus, $\bosigma^{\rmn}$ is estimated after the estimation of $\tbosigma$, for which $\bosigma^{\rmn} \succeq 0$ holds in low SNR scenarios.

Let us denote $\check{\hbfr}$ and $\check{\bfC}$, that refer, respectively, to $\hbfr$ and $\bfC$ without their elements corresponding to the diagonal of $\bfR$. The solution of (\ref{eq:min1}) w.r.t. $\tbosigma$ can then be obtained as
\begin{equation}
\begin{aligned}
  \htbosigma &= \argmin_{\tbosigma} \left\|\check{\hbfr} - \check{\bfC} \tbosigma \right\|^2_2 \\
  &\text{subject to~} \tbosigma \succeq 0, \left\| \tbosigma_q \right\|_0 = 1, q=1,\ldots, Q \text{,} 
\end{aligned}
\end{equation}
which is used in Algorithm 1.3. 

Afterward, the minimizer of (\ref{eq:min1}) w.r.t. $\bosigma^{\rmn}$ is given by
\begin{equation}
 \cbosigma^{\rmn} = \vectdiag\left(\hbfR-\hbfG \hbfRzero \hbfG^{\sfH} \right)  \text{.}
\end{equation}
Let us remove the bias introduced by the non-calibration sources as follows: we calculate the power
\begin{equation}
 \sigma_{r} = \frac{\bfa(\bfd_{r})^{\sfH} \left( \hbfR - \hbfG \hbfRzero \hbfG^{\sfH} \right) \bfa(\bfd_r)}{\left\|\bfa(\bfd_{r})\right\|_2^2 } \label{eq:sigman:1}
\end{equation}
of the residual sample covariance matrix for a random direction $\bfd_{r}$, where no source is supposed to be present. We then approximate $\bfa(\bfd_{r})^{\sfH} \bfa(\bfd_q) \approx 0$ for any $\bfd_{r} \ne \bfd_q$, which yields $\sigma_{r}$ as the sum of the sensor noise powers. By imposing $\sum_{p=1}^{P} [\bosigma^{\rmn}]_{p} = \sigma_{r}$, the unbiased solution is given by
\setlength{\arraycolsep}{0.0em}
\begin{equation}
 \hbosigma^{\rmn} = \check{\bosigma}^{\rmn} +\frac{1}{P} \left(\sigma_{r} - \bfun_{P \times 1}^{\sfT} \check{\bosigma}^{\rmn} \right) \bfun_{P \times 1} \text{,}
\label{eq:sigman:2}
\end{equation}
that concludes Algorithm 1.3.
\setlength{\arraycolsep}{5pt}

\section{Simulations}
\label{sec:simu}

The proposed method is tested using Monte-Carlo simulations and compared with the deterministic Cramér–Rao bound (CRB), that expresses a lower bound on the variance of any unbiased estimator. We test the algorithm in standard situations, with similar sensor locations, sky and parameters as in \cite{Wij2010, Wijthe, Ven2013}.

The antenna locations correspond to the LOFAR’s Initial Test Station \cite{Wij2004}, with $P = 60$ antennas disposed in a five-armed spiral. We assume a sky model at \SI{30}{MHz} ($\lambda = \SI{10}{m}$) consisting of $Q=2$ strong calibration sources and $Q^{\OUT}=8$ weak non-calibration sources, provided from the ten strongest sources in the table of \cite{Ben1962}. The total power of these sources is assumed to be 1\% of the total antenna noise power, a typical scenario for radio interferometers \cite{Wijthe}. Data are generated via the signal model given in (\ref{eq:signal}), in order to obtain the sample covariance matrix. We choose initially a coarse grid, with the same resolution for each coordinate and each calibrator. To avoid off-grid mismatch, we apply grid refinements \cite{Mal2005} until we achieve the theoretical limits given by the CRB.

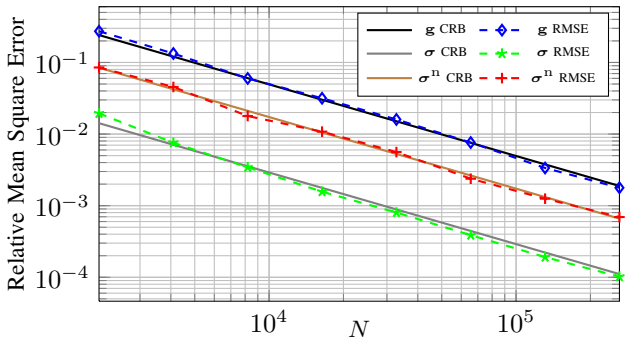
\begin{figure}[hb]
    \begin{tikzpicture}[]
   \begin{loglogaxis}[height=5.5cm, width=8.5cm, enlarge x limits=false, xlabel=$N$,ylabel= Relative Mean Square Error, legend entries={$\bfg$ CRB,$\bfg$ RMSE,
   $\bosigma$ CRB,$\bosigma$ RMSE,$\bosigma^{\rmn}$ CRB,$\bosigma^{\rmn}$ RMSE}, legend style={font=\tiny}, legend columns=2,
   label style={font=\small},grid=both, ticks=both,
      ylabel style={xshift=-0ex,yshift=-1ex},
   xlabel style={xshift=-0ex,yshift=+2.5ex},
   mark options={solid}]
    \addplot+[mark=none, draw = black, thick] table[x=N,y=crg] {resultats_2.txt};
    \addplot+[dashed, mark=diamond, draw = blue, thick] table[x=N,y=algog] {resultats_2.txt};
    \addplot+[mark=none, draw = gray, thick] table[x=N,y=crsigma] {resultats_2.txt};
    \addplot+[dashed, mark=star, draw = green, thick] table[x=N,y=algosigma] {resultats_2.txt};
    \addplot+[mark=none , draw = brown, thick] table[x=N,y=crnoise] {resultats_2.txt};
    \addplot+[dashed, mark=+, draw = red, thick] table[x=N,y=algonoise] {resultats_2.txt};
   \end{loglogaxis}
  \end{tikzpicture}
  \caption{Relative Mean Square Error on the undirectional antenna gains, the powers of the calibrators and the antenna noise powers. \label{fig:3}}
\end{figure}

\begin{figure}[t]
    \begin{tikzpicture}[]
   \begin{loglogaxis}[height=5.3cm, width=8.5cm, enlarge x limits=false, xlabel=$N$, ylabel= Mean Square Error, legend entries={$\bfd_1$ CRB,$\bfd_1$ MSE,$\bfd_2$ CRB, $\bfd_2$ MSE}, legend style={font=\tiny}, legend columns=2,
   label style={font=\small},grid=both, ticks=both,
   ylabel style={xshift=-0ex,yshift=-1ex},
   xlabel style={xshift=-0ex,yshift=+2.5ex},
   mark options={solid}]
    \addplot+[mark=none, draw = black, thick] table[x=N,y=crl1] {resultats.txt};
    \addplot+[dashed, mark=diamond, draw = blue, thick] table[x=N,y=algol1] {resultats.txt};
    \addplot+[mark=none, draw = gray ,thick] table[x=N,y=crl2] {resultats.txt};
    \addplot+[dashed, mark=star, draw = green, thick] table[x=N,y=algol2] {resultats.txt};
   \end{loglogaxis}
  \end{tikzpicture}
  \caption{Mean Square Error on the directions of the calibrators.\label{fig:1}}
\end{figure}
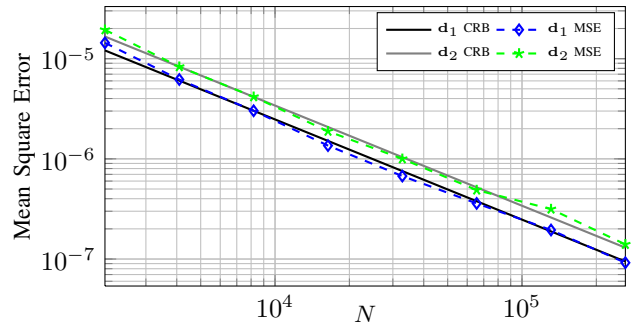

To investigate the algorithm performances, we perform 500 Monte-Carlo runs for each sample size. The variances of the errors on the complex undirectional gains, the calibrator directions and the powers, are plotted in Fig.\;\ref{fig:3} and Fig.\;\ref{fig:1}, as a function of the number of samples $N$ and compared to the corresponding CRB. As expected, the method appears to be asymptotically statistically efficient in low SNR scenarios, even with the presence of non-calibration sources. The (Relative) Mean Square Error may be slightly lower than the CRB for some values of $N$ as shown in Fig.\;\ref{fig:3} and Fig.\;\ref{fig:1}. This is mainly due to the fact that our algorithm takes into account the true/nominal direction of the calibrators (see (\ref{eq:fat_a})), whereas the classical CRB does not include this prior knowledge.

\section{Conclusion}
In this paper, we presented a novel iterative scheme for the calibration of radio interferometers, where different shifts affect the apparent directions and powers of the calibration sources. The proposed algorithm, named Iterative and Sparsity Based Calibration Algorithm (ISBCA), iteratively minimizes a weighting Least Squares function in order to estimate the complex undirectional antenna gains and their noise powers, whereas, it jointly estimates the directions and powers of the calibrators using the Normalized Iterative Hard Thresholding procedure. This leads to a statistically efficient, computationally efficient and robust scheme as shown by numerical simulations.

\bibliographystyle{ieeetr}
\bibliography{biblio}
\end{document}